\documentclass{article}

\usepackage{arxiv}

\usepackage[utf8]{inputenc} % allow utf-8 input
\usepackage[T1]{fontenc}    % use 8-bit T1 fonts
\usepackage{graphicx}
\usepackage{color}
\usepackage{epstopdf}
\usepackage{hyperref}       % hyperlinks
\usepackage{url}            % simple URL typesetting
\usepackage{booktabs}       % professional-quality tables
\usepackage{amsfonts}       % blackboard math symbols
\usepackage{nicefrac}       % compact symbols for 1/2, etc.
\usepackage{microtype}      % microtypography
\usepackage{lipsum}		% Can be removed after putting your text content
\usepackage{graphicx}
\usepackage{doi}
\usepackage{amsmath, amssymb, amsfonts}
\DeclareMathOperator*{\argmin}{arg\,min}
\title{A biophysical network model reveals the link between deficient inhibitory cognitive control and major neurotransmitter and neural connectivity hypotheses in schizophrenia}

\author{Konstantinos Spiliotis\\
	Institute of Mathematics\\ University of Rostock, D-18057\\ Rostock, Germany\\
	\texttt{konstantinos.spiliotis@uni-rostock.de} 
	\And
	{Giannis Kahramanoglou} \\
2nd Department of Psychiatry\\
National and Kapodistrian\\
University of Athens, Medical School\\ University General Hospital\\ ''ATTIKON'', 12462, Athens, Greece\\
	\texttt{i.kahra@gmail.com} 
		\And
	{Jens Starke} \\
	Institute of Mathematics\\ University of Rostock, D-18057\\ Rostock, Germany\\
	\texttt{jens.starke@uni-rostock.de} 
		\And
	{Nikolaos Smyrnis} \\
2nd Department of Psychiatry\\
National and Kapodistrian\\
University of Athens, Medical School\\ University General Hospital\\ ''ATTIKON'', 12462, Athens, Greece\\
	\texttt{smyrnis@med.uoa.gr
} 
		\And
	{Constantinos Siettos\thanks{Corresponding author: constantinos.siettos@unina.it}} \\
Dipartimento di Matematica \\
e Applicazioni “Renato Caccioppoli”\\ Università degli Studi di Napoli\\ Federico II, Napoli, Italy\\
	\texttt{constantinos.siettos@unina.it} 
}

\renewcommand{\shorttitle}{\textit{arXiv} Template}

\hypersetup{
pdftitle={A template for the arxiv style},
pdfsubject={q-bio.NC, q-bio.QM},
pdfauthor={David S.~Hippocampus, Elias D.~Striatum},
pdfkeywords={First keyword, Second keyword, More},
}

\begin{document}
\maketitle

\begin{abstract}
We address a biophysical network dynamical model to study how the modulation of dopamine (DA) activity and related N-methyl-d-aspartate (NMDA) glutamate receptor activity as well as the emerging Pre-Frontal Cortex (PFC) functional connectivity network (FCN) affect inhibitory cognitive function in schizophrenia in an antisaccade task. The values of the model parameters and the topology of the PFC-FCN were estimated by minimizing the differences between simulations and the observed distributions of reaction times (RT) during the performance of the antisaccade task in 30 patients with schizophrenia and 30 healthy controls. We show that the proposed model approximates remarkably well the predicted prefrontal cortical DA hypo-activity and the related NMDA receptor hypo-function as well as the FCN dysconnection pattern that are considered as the major etio-pathological  hypotheses to explain cognitive deficits in schizophrenia. 
\end{abstract}

% keywords can be removed

\section{Introduction}

Despite major advancements in contemporary neuroscience research, the mechanisms that underpin the pathophysiology of major psychiatric disorders such as schizophrenia remain elusive. A major challenge revolves around the development of an integrated framework, which can bridge the scales where the questions are asked and the answers are required: the microscopic level (molecular genetics, neurophysiology and neuropharmacology of the neuron) and the macroscopic level where one observes and measures changes in the behavioral and cognitive functions combined with alterations in the neurophysiology and anatomical-functional imaging of the whole brain \cite{Hig07, Cann15,Fat09,Fris16,Kub07,Ruiz13,Kel18}. Between these two scales, it extends the mesoscopic scale, where one attempts to describe the activation and communication of specific neuronal networks located in different  brain areas. The coordinated activity of these networks gives rise to emerging/macroscopically-experimentally observed behavior and corresponding global brain activation patterns. The mesoscopic level of description has been the focus of many studies on the electrophysiology of rodents and primates using single neuron, and more recently multi-neuron recording techniques, combined with histology and more recently optogenetics \cite{Jen15,Krav12}. \par
\begin{figure}%[tbhp]
\centering
\includegraphics[width=1.\linewidth]{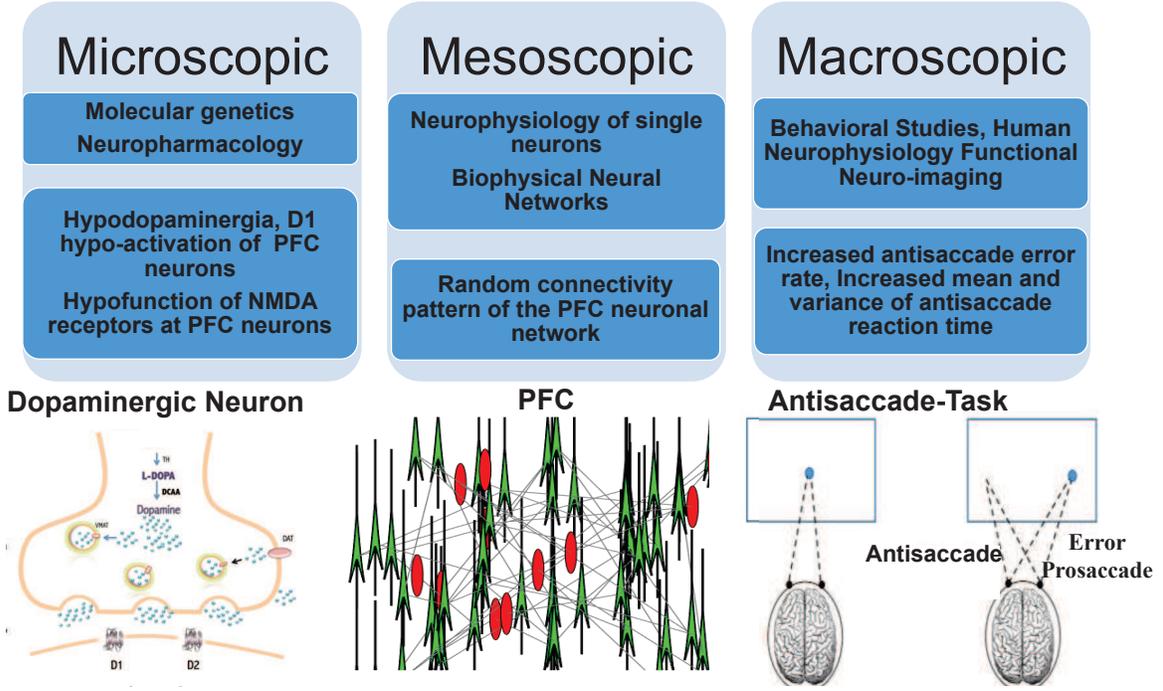}
\caption{Schematic of the Experiment and different levels of inquiry ranging from the microscopic level: molecular genetics, neurophysiology and neuropharmacology of the neuron, to the mesoscopic level: activation and communication of specific networks of connected neurons located in different areas of the brain, for example of the PFC network (green triangles and red elipses are for pyramidal excitatory neurons and interneurons respectively), and finally to the macroscopic level: changes in the behavior and cognitive functions combined with changes in the neurophysiology and anatomical-functional imaging of the whole brain. Behavioral experiment: antisaccade task. Starting with a fixated cue centered on a screen, a peripheral stimulus (S) appears, the subject is instructed to  look at the opposite direction (antisaccade) suppressing the saccadic motion (error prosaccade) exerting an inhibitory control over the natural tendency to look towards the visual stimulus.}
\label{fig:experiment}
\end{figure}

One way to shed light on the mesoscopic organization and functionality is through the use of detailed biophysical neuronal network models. In general, the biophysical modelling approach is described by the following steps (see Fig.~\ref{fig:experiment}): a) identification of a specific behavioral/cognitive function that is deviant in schizophrenia, b) construction of a detailed biophysical neuronal model based on the knowledge acquired from detailed experiments and/or from animal studies including the identification of the neuronal  networks at specific brain areas that govern the corresponding behavioral/cognitive function, c) use of the biophysical network model to study structural and functional connectivity changes as predicted by existing neuropharmacological and sub-neuronal function hypotheses, d) test how these changes have an impact on the predicted changes in the cognitive function, and finally, e) validation of the results of the neuronal network model using experimental data to test whether the presence of the predicted changes of the network model result in the same cognitive/behavioral deviance as observed e.g. in schizophrenia.\par
At the microscopic level, the dopamine hypothesis is a prominent theoretical framework that relates the mechanism of the antispychotic medications to psychopathology in schizophrenia \cite{Cann15, How09, Bris14}. The hypothesis suggests that hyper-dopaminergia or hyperactivity at D2-dopamine receptors in subcortical limbic system structures leads to the appearance of positive symptoms of the disorder (delusions, hallucinations) and that the hypo-dopaminergia or hypoactivity of D1-dopamine receptors in the prefrontal cortex leads to the appearance of negative symptoms and cognitive dysfunctions \cite{How09, Bris14}. 
Another prominent hypothesis, the glutamate hypothesis is linked with the functional modulation of N-methyl-D-aspartate (NMDA) receptors \cite{Fat09, Cann15, Lew09, Gon12, Sny13}. NMDA receptors which are located in cortical pyramidal neurons are activated by glutamate, the main excitatory neurotransmitter in the mammalian brain. Activation of pyramidal neurons in the deep cortical layer results in the activation of GABAergic inhibitory interneurons in the cortex forming a local functional network \cite{Lew09, Cann15, Sny13}. The glutamate hypothesis states that a hypoactivity at NMDA receptors results in an imbalance of this excitation-inhibition network. Substances that are NMDA antagonists, such as phencyclidine (PCP) and ketamine can produce both positive symptoms and cognitive dysfunctions in healthy individuals that resemble schizophrenia, as well as worsening of such symptoms in schizophrenic patients \cite{Laht14, Sny13}. Furthermore, postmortem studies in patients have shown significant changes in NMDA receptor protein expression in the prefrontal cortex \cite{Sny13, KRIS07}.\par
Remarkably, these two phenomenological independent hypotheses might share the same pathophysiological pathway in the prefrontal cortex. The D1 receptor signaling in cortical neurons is regulated by the action of NMDA receptors and vice versa \cite{Scott02, Che04, How09}. Activation of D1 receptors in the prefrontal cortical pyramidal neurons by the agonist SKF81297 lead to an increase of the steady-state NMDA evoked current \cite{Che04} and this effect is canceled by incubation of neurons in $[\text{Ca}^{2+}]$ free medium. Inversely, activation of NMDA receptors by glutamate results in the recruitment of D1 receptors in cortical neurons while they don't have any effect on the distribution of D2 receptors \cite{Scott02}. The dysconnection hypothesis describes the above synaptic complex: it suggests that psychosis  in terms of aberrant neuromodulation of synaptic efficacy mediates the influence of intrinsic and extrinsic (long-range) connectivity \cite{Fris16}. It suggests that the key pathophysiology lies in the interactions between NMDA receptor functions, the modulatory effect of neurotransmittion and the mediated changes in synaptic efficacy. The NMDAR-mediated plasticity affects  the functionality at the level of neuronal circuits which in turn leads to an abnormal functional integration (dysconnection) of cognitive functions among brain regions in schizophrenia.\par
Here, the cognitive function that we study is the inhibition of the response in healthy controls and patients with schizophrenia while performing an antisaccade task. In the particular task, subjects are instructed to look at the opposite direction of a visually presented stimulus, thus exerting inhibitory control over the natural tendency to look towards the visual stimulus \cite{Hall78}. There is a large body of literature that confirms a deficit in this task in patients with schizophrenia compared to healthy controls \cite{Pan19}. More specifically, it has been reported that patients produce more erroneous responses towards the visual target than healthy controls while the response latency for correct antisaccades for these patients is longer and more variable than that observed for healthy controls \cite{Bas08, Pan19}. A large body of animal and human studies has confirmed the critical role of the prefrontal cortical areas such as the frontal eye field, the supplementary eye field and especially the dorsolateral prefrontal cortex in the performance of the antisaccade task \cite{Mun04, Muri05, Brown07}. There is also evidence that deficits in the performance of the antisaccade task in schizophrenia may be related with hypo-activity in specific areas of the prefrontal cortex that are activated during this task such as the dorsolateral prefrontal cortex \cite{Mun04, Coe17,Pan19}, (see also Fig.~\ref{fig:experiment}).\par
Here, building up on previous work \cite{Dur03,Cut07a, Cut07b, Kahr08}, we attempt to bridge the microscopic and macroscopic levels by addressing a multiscale neuronal network model of two brain regions, namely the prefrontal cortex (PFC) and the superior colliculus (SC) for the simulation of the antisaccade task. The model incorporates competing neuronal activity at the intermediate layer of the SC that is driven both by a planned input (the antisaccade command) and a reactive input (the erroneous prosaccade command). The output of the model are the reaction time distributions for the correct antisaccades  and erroneous prosaccades, as well as the percentage of the error prosaccades.
The PFC activity is approximated by leaky spiking integrate and fire pyramidal neuronal dynamics as described in \cite{Dur03}, while the connectivity of the PFC neurons is allowed to vary from a ring-like network to a completely random topology. Unlike previous modeling attempts \cite{Cut07a, Cut07b,Cut14}, we did not assume any a-priori knowledge of the biophysical parameter values and the PFC neuronal connectivity structure; these are estimated in a strict numerical way, by ``wrapping" around the detailed network model a numerical optimization structure, namely the Levenberg-Marquardt algorithm \cite{web:lm} to fit the experimental behavioral data. Then, we used the calibrated model to bridge microscopic and macroscopic scales, namely to study how changes in the dopamine levels acting on D1 receptors and the related hypo-activity of NMDA receptors on simulated prefrontal cortical neurons (microscopic level) give rise to antisaccade performance differences in patients with schizophrenia vs. healthy controls. In addition, the fact that the detailed connectivity of the biophysical neuronal network model of the PFC was also allowed to vary, gave us the opportunity to (qualitatively) study how changes at the microscopic level can give rise to changes at the mesoscopic level (namely how the topology of the specific network changes) that in turn give rise to macroscopic changes at the experimentally observed behavioral performance. Our hypothesis is that decreased DA levels acting on D1 receptors associated with hypofunction of NMDA receptors in the PFC would predict antisaccade performance differences as observed in schizophrenia, namely a large increase in the error rate as well as an increase in the mean and variance of the reaction times for correct antisaccades. Furthermore, in accordance to the dysconnection hypothesis, we predicted that the deviances in the microscopic level would mediate the macroscopic effects via a change at the mesoscopic level, namely a change in the functional connectivity of the underlying neuronal network. 
%Another new insights (at least to our knowledge) is our findings that
%the Indeed, our results confirm the above hypotheses, thus showing that
%the emergent connectivity of the PFC network was more random in patients with schizophrenia compared to healthy controls  (see Fig.~\ref{fig:experiment}).
%In this work we present an application of this method to the study of cognitive deviance in schizophrenia. We will show how a detailed biophysical neural network model combines specific deviances in the microscopic level as predicted by two prominent hypotheses based on the neuropsychopharmacology of schizophrenia to give rise to specific functional deviances observed in a very well-studied cognitive dysfunction paradigm in schizophrenia. Moreover we will show the emergence of model structure deviances at the mesoscopic (neural network) level that lead to testable hypotheses for future studies. 

\section{Material and Methods}
\subsection*{The Prefrontal Cortex network model}

\begin{figure}%[tbhp]
\begin{center}
\begin{picture}(400,240)
\includegraphics[width=12.5cm]{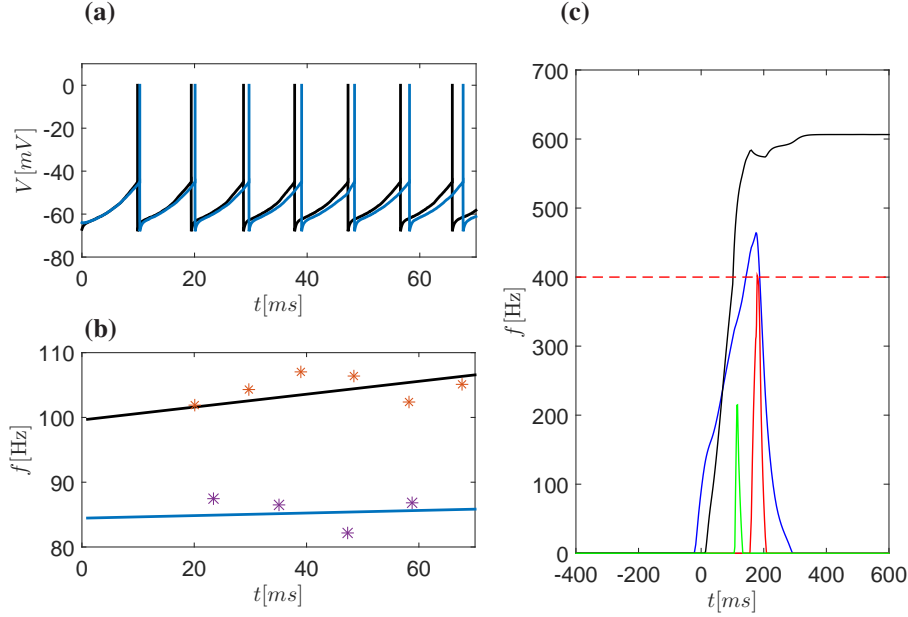}% This is a *.eps file
\put(-320,230){\textbf{(a)}}
\put(-140,230){\textbf{(c)}}
\put(-320,110){\textbf{(b)}}
\end{picture}
\end{center}
\caption{Simulation of  neurons in PFC and SC. 
\textbf{(a)} Membrane potentials for two neurons in the PFC network. One of these neurons fires with higher frequency (black line). \textbf{(b)} The corresponding linear rising phase (see sec. \ref{sec:SC}) for the previous PFC neurons. A neuron fires with slightly higher frequency  (black line) giving a bigger slope in the linear frequency-time approximation. \textbf{(c)} Firing rates in SC of buildup and burst neurons from a simulation run of the antisaccade task. Buildup neurons (black line and blue lines) encoding reactive and planned saccade tasks. When the activity of buildup neuron crosses a threshold (400HZ red dash line)  an erroneous prosaccade (light green) is initiated followed by a correct saccade.}
\label{fig:Simulations}
\end{figure}

We build up on a model proposed in \cite{Dur03} using spiking integrate and fire neurons whose dynamics are governed by the current balance equation:
\begin{equation}
C\frac{dV_i}{dt}=-I_\text{{LEAK}}-I_\text{{AHP}}-I_\text{ADP}-I_\text{NMDA}-I_\text{GABA}+I_\text{INJ},
\end{equation}
where $C$ is the membrane capacity, $V_i$ is the membrane potential of the $i-$th neuron, $I_\text{LEAK}$  $I_\text{INJ}$, $I_\text{AHP}$ and $I_\text{ADP}$ are the leak, injected, after hyper polarizing and depolarizing currents respectively; $I_\text{NMDA}, I_\text{GABA}$ correspond to the excitatory and inhibitory synaptic currents, respectively.\\
If $V_i>V_\text{thres}$, a spike is being fired and the potential returns to $V_i=E_\text{LEAK}$. The leak current is given by:
\begin{equation}
I_\text{LEAK}=g_\text{LEAK}(V_i-E_\text{LEAK}),
\end{equation}
where $g_\text{LEAK}$ is the corresponding conductance. The after hyperpolarizing current is given by:
\begin{equation}
I_\text{AHP}=g_\text{AHP}e^{-(t-t_{sp})/\tau_\text{AHP}}(V_i-E_\text{AHP}),
\end{equation}
where $t_\text{sp}$ is the time of the last spiking and $\tau_\text{AHP}$ reflects  the decay time scale of the current; $g_{AHP}$ is the conductance and $E_\text{AHP}$ is the reverse potential.\\
The after depolarizing current, $I_\text{ADP}$ is due to the $ [\text{Ca}^{+2}]_i$ influx in pyramidal neurons and is modeled according to the following equations \cite{Dur03}:
\begin{equation}
\begin{cases}
 I_\text{ADP}&=g_\text{ADP}m_{i}(V_i-E_\text{ADP}) \\
\frac{dm_i}{dt}&=\frac{(m_{\infty}-m_i)}{\tau_\text{ADP}}\\\
 [\text{Ca}^{+2}]&=A_\text{{Ca}}\sum_{t_{sp}}e^{-(t-t_{sp})/t_{off}}-e^{-(t-t_{sp})/t_{on}},
\end{cases}
\end{equation}
where $m$ is the gating activation variable related to $[\text{Ca}^{+2}]$ influx, as:
\begin{equation}
m_{\infty}= (1+exp(\gamma(\theta-[\text{Ca}^{+2}]_i))^{-1}.
\end{equation}
$t_{\text{on}}, t_{\text{off}}$ are constant time scales marking the onset and offset of the $[\text{Ca}^{+2}]$ influx. $E_\text{{ADP}}$ is the reverse potential, $g_{\text{ADP}}$ is the conductance. The parameters $\gamma$ and $\theta$  control  the sharpness of the equilibrium function $m_{\infty}$, while $A_{\text{Ca}}$ controls the  calcium concentration resulting from the spike sequence \cite{Dur03}.
%\begin{figure}%[tbhp]
%\centering
%\includegraphics[width=1.\linewidth]{gen2.eps}
%\caption{Computational model consists of two part, the PFC which is modeled as small world network with $N=100$ neurons. Red circles in PFC denote excitatory neurons while black squares correspond to inhibitory ones. Staring with a ring structure  with $2k$ local neighbors for each node. With probability $p$,a remote edge replaces each local connection. 
%The Superior Colliculus (Tectum) contains 31 neurons 15 burst (red), 15 built up(green) and 1 fixation (blue), all to all connected.
%\textbf{
%Costas: I am now convinced that the figures with the small-world networks should be deleted. They will create just problems to someone non-mathematician. A ring network is not so realistic, on the other hand to create a realistic network for the interconnections of PFC is almost impossible as few things are known among them that the network is a small-world network... We needed this to be able use the rewiring probasbility p as optimization parameter as this would be impossible for an arbitrary small world network...}
%\label{fig:Network}
%\end{figure}
In our model, the neurons are connected through a Watts and Strogatz (WS) small-world topology \cite{Wat98, Bull09, Stam07}. The network is reflected on the synaptic currents defined by the activation variable $s$ which for each single neuron is given by \cite{Lai02,Erm12,Com00}:
\begin{equation}
\frac{ds_i}{dt}=\alpha(1-s)H(V_i-\theta_{0})-\beta s_i,
\end{equation}
where $H(V)$ is the step function.\\
If $V_i<\theta_{0} \implies H(V_i-\theta_{0})=0 $ and the synaptic equation takes the form: 
\begin{equation}
\frac{ds_i}{dt} =  - \beta s_i \Rightarrow s_i = c{e^{ - \beta t}} \Rightarrow s\to 0, t \to +\infty. 
\end{equation}
Hence, the synapse turns off with $\beta$ as time scale.\\
Instead, if $V_i>\theta_{0} \implies H(V_i-\theta_{0})=1 $ and the synaptic equation takes the form:
\begin{equation}
\frac{ds_i}{dt}= \alpha \left( {1 - s_i} \right) - \beta s_i.
\end{equation}
The above has the solution 
\begin{equation}
s_i(t) = c\frac{{{e^{ - (\alpha  + \beta )t}}}}{{\alpha  + \beta }} + \frac{\alpha }{{\alpha  + \beta }}.
\end{equation}
The above expression implies that synaptic jumps are given by $\frac{\alpha{}}{\alpha+\beta}$.\\
Usually, fast synapses have $\alpha ,\beta  = \text{O}\left( 1 \right)$, while for slow synapses  $\alpha=\text{O}\left(1 \right)$ and $\beta  = \text{O}\left(\epsilon \right)$ meaning that fast synapses activate and deactivate on fast time scales, while the slow synapses activate on fast and deactivate on slow time scales \cite{Lai02,Erm12}. The form of the excitatory and inhibitory synaptic currents are given, respectively, by:
\begin{equation}
I_{\text{NMDA}}=g_{\text{NMDA}}(V_i-E_{\text{NMDA}})\sum{s_{j}}, 
\end{equation}
for $E_{\text{NMDA}}=0$, and
\begin{equation}
I_{\text{GABA}}=g_{\text{GABA}}(V_i-E_{\text{GABA}})\sum{s_{j}}, 
\end{equation}
for $E_{GABA}=-80$.
The summation is taken over the presynaptic neurons.\\
Excitatory currents result from pyramidal neurons, while inhibitory currents result from inter-neurons. Out of the $N$ neurons, $\frac{3N}{4}$ are pyramidal neurons, while the rest are inter-neurons. Fig.\ref{fig:Simulations}(a) shows the dynamics of two neurons in the  PFC network. The neurons although start with nearby initial conditions they exhibit slightly different firing rates (see Fig.\ref{fig:Simulations}(b)).

\subsection*{The Superior Colliculus (Tectum) model}
\label{sec:SC}
The neuronal model is a classic on-center off-surround leaky competitive integrator \cite{Tra01,Tay99,Ara94,Kop95}. The internal state which represents the firing rate $x_i(t)$  of the $i-$th node is governed by the following differential equation:
\begin{equation}
\tau \frac{dx_i}{dt}=-x_i+\sum_{j}{A_j w_{ij}+I_p+I_r-u_0+I_n},
\end{equation}
where $\textbf{w}=(w_{ij}), i, j=1,2,...N$ is the synaptic efficacy from neuron $i$ to neuron $j$, $A$ is the activity function of node $j$, $Ir$ and $Ip$ are the reactive and planned inputs, respectively, that the Tectum receives from other cortical areas, $u_{0}$ is a global inhibition term, and $I_{n}$ is the background noise.\par
The value of $u_{0}$ is set to zero for the buildup neurons and to a large value for the burst neurons as burst neurons are shown to have a discharge activity only after the activity of the buildup neurons reaches a certain threshold \cite{Mun95}. The activity of the burst neurons is restored back to zero only when these neurons surpass an activity level equal to 80\% of their theoretical maximum discharge rate \cite{Tra01,Cut07a}.
The activity function $A_j(t)$ of a neuron $j, j=1, 2,..N$ representing the average membrane potential is given by a sigmoid function
\begin{equation}
 A_i(t)=\frac{1}{1+e^{-bx_i+\theta}},
\end{equation}
where $b$ is the steepness and $\theta$ is the offset of the sigmoid function.
The interaction matrix $w$ allows for lateral interactions between neurons in the same colliculus and between neurons located in opposite colliculi sites \cite{Mun98,Mer98}; it depends only on the spatial distance and it is positive (excitatory) i.e. $w_{ij}>0$ for short distances, and negative(inhibitory) i.e. $w_{ij}<0$ for long distancing neurons. Thus, the interaction matrix is described by
\begin{equation}
 w_{ij}=a\frac{e^{-(j-i)^2}}{2\sigma_a^2}-b\frac{e^{-(j-i)^2}}{2\sigma_b^2}-c,
\end{equation}
where $a$, $b$, and $c$ are free parameters and $\sigma_a$ and $\sigma_b$ are spatial parameters.\par
Two competing input signals are integrated in the Tectum: a planned and a reactive. In the model, the origins of these two input signals differ: the reactive signal is considered to originate from the Post Parietal Cortex (PPC), whereas the planned signal from the frontal executive centers of the brain (the PFC).
The reactive input signal is governed by a simple differential equation reading:
\begin{equation}
\begin{cases}
\frac{dI_r}{dt}=A\mid slope_r\mid, if \ t \geq t_{on}+t_r^{delay} \ and \ I_r\leq I_r^{max} \\
\frac{dI_r}{dt}=-a_rA \cdot I_r, if \ t \geq t_{on}+t_r^{delay} and \ I_r> I_r^{max} \\
\frac{dI_r}{dt}=-a_r A \cdot I_r, else\\
\end{cases}
\end{equation}
$a_r$ are the integration strengths, $I_r^{max}$ is a theoretical maximum allowed activity for the reactive input, $t_{on}$ is a constant indicating the onset of the incoming signal $I_r$. $A$ is the strength of the rising phase and $slope_r$ represents the linear rising phase of the reactive input and follows a bell like probability distribution function (normal with mean value $\overline{x}=5.27$ and standard deviation $std=2$) \cite{Cut07a,Cut07b}. The reactive input reflects the sensory information reaching the Tectum without extensive information processing and it is taken to follow closely the onset of a visual stimulus in the periphery with a delay, $t_r^{delay}$. \par
The planned input signal has also a linear rising phase \cite{Red00} before it reaches its theoretical maximum value and is governed by:
\begin{equation}
\begin{cases}
I_p=A\mid slope_p\mid, if \ t \geq t_{on}+t_p^{delay} \ and \ I_p\leq I_p^{max} \\
I_p=A \cdot I_p^{max}, if \ t \geq t_{on}+t_p^{delay} and \ I_r> I_p^{max} \\
I_p=0, else. \\
\end{cases}
\end{equation}
$I_{p}^{max}$ is the theoretical maximum activity of the planned input, $A$ is the strength of the planned input and $slope_p$ is the slope of the planned input rising phase and it is extracted from the neural integration (time evolution) of the PFC according to the following way: for each PFC neuron, we perform a linear approximation of the corresponding firing rate (which is called linear rising phase and it is calculated in the first 70 msec). The mean value of these slopes defines $slope_p$. The planned input reflects the processing of the planned antisaccade by higher processing centers such as the frontal eye fields (FEF), the supplementary eye fields (SEF), and the dorsolateral prefrontal cortex (DLPFC) to determine the behavioral response that would be appropriate for the given task instruction and it is considered to take longer, i.e. $ t_p^{delay}>t_r^{delay}$ for processing than the reactive input due to additional cortical processing. \par
The strength of both planned and reactive signals is given by:
\begin{equation}
 A(i,j)=A_0 e^{\frac{-(j-i)^2}{2\sigma_A^2}},
\end{equation}
where $i$ and $j$  are the indices of nodes and $\sigma_A$ is the standard deviation of the Gaussian.
%The width of the Gaussian was derived from the shape of movement fields of saccade-related neurons in the monkey Tectum \cite{Mun2,Tra1}.
%In the  simulations, the threshold level (horizontal line at about 400 Hz) was appropriately adjusted, so that both buildup nodes encoding the reactive and planned inputs crossed the threshold and an erroneous prosaccade (error burst) was initiated followed by a correct saccade (correct burst). The aSRT is defined  as the time when a randomly selected SC burst neuron $x_i$ reaches the maximum value\cite{Cut1,Cut2}.
The width of the Gaussian function was derived from the shape of movement fields of saccade related neurons in the monkey SC \cite{Mun95, Tra01}. In the simulations, the threshold level (horizontal line at about 400 Hz) was appropriately adjusted, so that both buildup neurons encoding the reactive and planned inputs cross the threshold and an erroneous prosaccade (error burst) was initiated followed by a correct saccade (correct burst). We considered a correct antisaccade response, if the movement of the eyes went in the opposite direction of the stimulus; all other cases were considered to be error prosaccades. Saccade reaction times were estimated to be the time interval from the onset of peripheral stimulus until the time the activity of the burst neurons deviated from zero plus 20 ms (approximate time required for burst neuron signal to reach the eye muscles) \cite{Cut07a,Cut07b}. Fig.\ref{fig:Simulations}(c) shows the firing rates in SC of two buildup and two burst neurons from a simulation run of the antisaccade task.

\section{Experimental Data Analysis}
\label{sec:Experimental}
\subsection*{Experimental Data}
The experimental data used in this study are derived from a previous study of the group \cite{Smy09}. Briefly the antisaccade performance data of 30 male patients (age span 18-30 years) with DSM-IV schizophrenia formed the patient group. Patients were evaluated at the Psychosis Unit of the Psychiatry Department of the National and Kapodistrian University of Athens at Eginition Hospital and the diagnosis of schizophrenia was confirmed by a trained psychiatrist with the use of the Mini International Neuropyschiatric Interview (M.I.N.I., version 5.0.0., DSM IV) \cite{She97}. Exclusion criteria consisted of the following: neurological disorder (epilepsy, multiple sclerosis etc), mental retardation and drug abuse within the last year before evaluation. All patient participants were receiving antipsychotic medication and were in a stable phase of the disorder during testing. 
%\begin{table}%[tbhp]
%\centering
%\caption{Parameters of the PFC model.\textbf{Costas: (1) the table is too long and narrow. We should use four columns here. (2)There are also missing values for the parameters. (3) Units are still missing. (4) We should put a reference next to each value.. (5) In the caption we say about the parameters for the PFC and not for the SC.... we hould put them together here....}}
%\begin{tabular}{|l|r|}
%\hline
%Name of the parameter &  value  \\
%\midrule
%$E_{LEAK}$ & -68 \\
%$E_{NMDA}$ & 0  \\
%$E_{GABBA}$ & Eleak-5 \\
%$E_{AHP} $ & Eleak \\
%$E_{ADP}$ & 10  \\
%$g_{ADP}$ & 0.078 \\
%$g_{AHP}$ & $N(r1,r2)$\\
%$g_{ABBA}$ & 0.07  \\
%$g_{NMDA}$ & r3*0.02 \\
%$g_{LEAK}$ & 0.012 \\
%$A_{Ca}$ & 0.08\\
%$\gamma$ & 0.09 \\
%$\tau_{AHP} $ & 5 \\
%$\tau_{ADP} $ & 0.68\\

%$t_{on}$ & 1 \\
%$t_{off}$ & 220 \\
%$\theta$ & 0.07  \\
%$\alpha$ & 1 \\
%$\beta$ & 0.05 \\
%$\theta_0$ & -55  \\
%$V_{thres}$ & -45  \\
%$k$ & 4  \\
%$I_{inj}$ & 8\\
%$r_1$ &   \\
%$r_2$ &   \\
%$r_3$ &   \\
%$p$ &  \\
%\bottomrule
%\end{tabular}
%\end{table}
The control sample was derived from the ASPIS (Athens Study for Psychosis proneness and Incidence of Schizophrenia) data base \cite{Smy02,Smy03,Evd02}. For the purposes of ASPIS oculomotor task data (smooth eye-pursuit, saccade, antisaccade, visual fixation) were collected from a population of 2120 conscripts of the Greek Air Force aged 18-25 years. Valid antisaccade data were obtained from 2006 individuals.\par 
Oculomotor tasks were performed in a set up that has been described in detail in our previous studies \cite{Smy02,Smy03,Evd02}. The antisaccade task was preceded by a calibration procedure, with saccades at 5 and 10 degrees to the left and to the right of a central fixation point. Each antisaccade trial started with the appearance of a central fixation stimulus (white cross 0.3 x 0.3 degrees of visual angle). After a variable period of 1-2 sec, the central stimulus disappeared and a peripheral stimulus (same white cross) appeared randomly at one of the 9 prescribed distances (2-10 degrees at 1 degree intervals) either to the left or to the right of the central fixation stimulus. The subjects were instructed to make an eye movement to the opposite direction from that of the peripheral stimulus as quickly as possible. Each subject performed 90 trials. \par
An interactive PC program (created using the TestPoint CEC) was used for the detection and measurement of saccades from the eye movement record \cite{Smy02,Smy03,Evd02}. We excluded trials with artifacts (blinks, etc.) in the period extending from 100ms before the appearance of the peripheral target to the end of the first saccade as well as trials for which an eye movement occurred within the 100-ms period before the appearance of the peripheral target. In addition, in order to avoid including predictive movements or too slow responses, we excluded trials with reaction times that were not within the window of 80-600ms. Using the above criteria, the minimum number of valid trials retained for each subject was 30 and the maximum was 90. 
\begin{figure}[t]
\begin{center}
\begin{picture}(400,190)
\includegraphics[width=12.5cm]{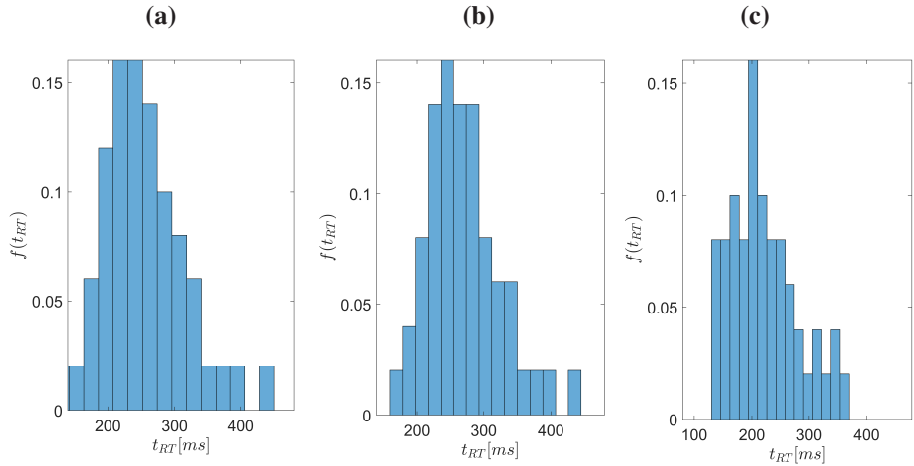}% This is a *.eps file
\put(-300,170){\textbf{(a)}}
\put(-180,170){\textbf{(b)}}
\put(-75,170){\textbf{(c)}}
\end{picture}
\end{center}
\caption{Healthy controls. Distributions of reaction times (RT) during the experiment (see sec.~\ \ref{sec:Experimental}). \textbf{(a)} Distribution of RT which includes correct antisaccades and wrong prosaccades; the main statistical moments are found to be: mean value $\bar{t}_{RT}=253$, standard deviation $s=60.83$. \textbf{(b)} Distribution of antisccade only RT:  mean value $\bar{t}_{RT}=270.84$ and standard deviation $s=55.53$. \textbf{(c)} Distribution of error prosaccases RT: mean value $\bar{t}_{RT}=208.94$ and standard deviation $s=42.59$.}
\label{fig:CON_EXP}            
\end{figure}

\begin{figure}[t]
\begin{center}
\begin{picture}(400,200)
\includegraphics[width=12.5cm]{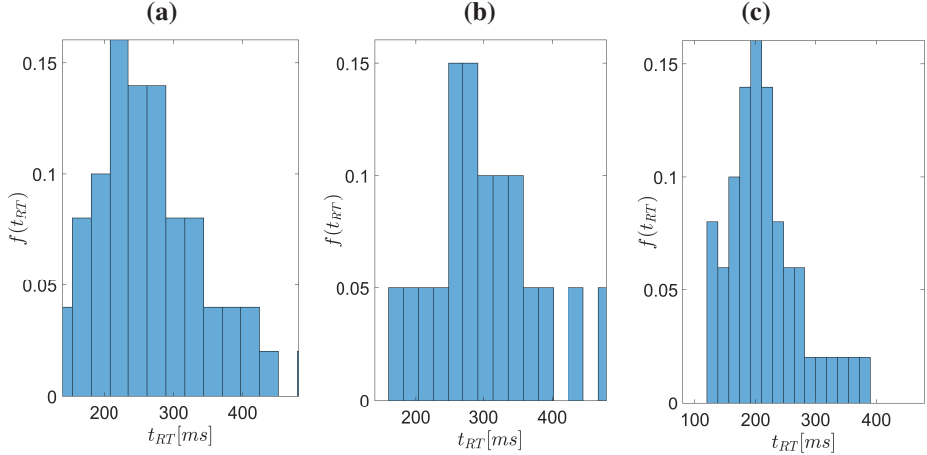}% This is a *.eps file
\put(-300,170){\textbf{(a)}}
\put(-180,170){\textbf{(b)}}
\put(-75,170){\textbf{(c)}}
\end{picture}
\end{center}
\caption{Distributions of reactions times (RT) of schizophrenic patients during the experiment (see sec.\ \ref{sec:Experimental}). \textbf{(a)} Distribution of RT which includes correct antisaccades and wrong prosaccades: mean value $\bar{t}_{RT}=265$, standard deviation $s=81$. \textbf{(b)} Distribution of antisccade only RT: $\bar{t}_{RT}=304$, standard deviation $s=77$. \textbf{(c)} Distribution of error prosaccases RT: mean value $\bar{t}_{RT}=215$, standard deviation $s=60$.}
\label{fig:SZ_exp}            
\end{figure}

\subsection*{Data Analysis}
The error rate was computed for all patients and healthy controls as the percentage of erroneous prosaccades towards the peripheral target over all valid trials. The mean value and standard deviation for each group was then computed and compared with the error rate of the simulation data using the t-test at a 0.05 level of significance. \par
The reaction times for valid trials for each subject were used to derive the cumulative reaction time distribution for each subject. Then, the reaction times for correct antisaccades and erroneous prosaccades were processed separately for each subject to derive the corresponding cumulative reaction time (RT) distributions. For all RT distributions, we also calculated the percentiles (from 5\% to 95\% with a step of 5\%). The percentiles were then averaged across the group to give the average group percentiles  that are plotted in the average cumulative distribution. Ratcliff \cite{Ratc79} showed that this average distribution retains the basic shape characteristics of the individual distributions. Thus, we created experimental averages of the RT distributions for (a) all RT, (b) correct antisaccade RT only, and (c) error prosaccade RT.  Fig.~\ref{fig:CON_EXP} depicts the resulting experimental RT distributions for the healthy controls, while Fig.~\ref{fig:SZ_exp} illustrates the RT distributions for the patients. The correct antisaccade cumulative RT distributions for each group were then compared with the corresponding average cumulative RT distributions derived from the model simulation using the Kolmogoroff-Smirnov test at a 0.05 significance level. 

\subsection*{Numerical Optimization}
Our PFC neuronal network consisted of $N_1 = 75$ pyramidal neurons and $N_2 = 25$ inter-neurons. The SC neuronal network had $N_3 = 31$ neurons (15 buildup neurons, 15 burst neurons and 1 fixation neuron). Ensembles of the PFC network topology were constructed using the WS algorithm \cite{Wat98, Stam07, Bull09}, thus allowing the connectivity to vary between ring and random regular structures; the connectivity topology is adjusted by the switching probability $p$.
In line with biological theories, variations in the receptors functionality (DA, NMDA) as well as alterations in the PFC network topology are reflected in the conductances of the  synaptic currents $I_{\text{NMDA}}, I_{\text{GABA}}$ as well as in the ionic currents e.g. calcium-activated potassium $I_{\text{AHP}}$ \cite{Malenka86}. Here, we investigated these alterations in the conductances of currents $I_{\text{NMDA}}$ and $I_{\text{AHP}}$ by extracting the conductance $g_{\text{AHP}}$ from a normal distribution with mean value $r_1$ and standard deviation $r_2$, while the $g_{\text{NMDA}}$ was our third parameter $g_{\text{NMDA}} = r_3$ to be tuned.\par
For each set of the model parameters $\textbf{r} = (r_1, r_2, r_3, p)$, the dynamic model produces a different PFC activity (defining a distribution of trials), thus resulting to different values of the $slope_p$. The resulting distributed values of $slope_p$ over all trials are then fed into the SC model in order to produce the reaction times for correct antisaccades as well error prosaccades RT. The number of simulation trials was set to $N = 100$. \par
The values of the model parameters $r_i ,i=1,2,3$, as well as the switching probability $p$ for the construction of the network topology were estimated numerically by minimizing the residual of the objective function defined as the norm difference of the distribution functions $f(\cdot)$ of the RT (both antsaccade and error prosaccade) between experimental and those resulting from simulations, i.e.: 
\begin{equation}
\argmin_{r\in \mathbb{R}_{+}^{4}} obj(\textbf{r}):=
\left\lbrace r: min|| {f(x_i,\textbf{r})}_{model}- \ {f(x_i)}_{Data} ||,\ \textbf{r} \in \mathbb{R}_{+}^{4} \right \rbrace 
\end{equation}
where the norm is the Euclidean distance between the $n$ points of approximate and experimental RT distributions functions. The procedure was implemented separately in the two groups of healthy controls and patients.
The minimization problem was solved using the matlab implementation of the Levenberg-Marquardt algorithm (function nlqnonlin \cite{matlab})\cite{web:lm}.
%Nelder-Mead optimization method \cite{Kel99,Lag98}. The Nelder–Mead method endeavors to minimize a real function $f:\mathbb{R}^n \rightarrow \mathbb{R}$ using only function values, without taking into account derivative information\cite{lagaria}.
The step size tolerance was set to $\text{tol}(X)=0.001$ 
%( $||x_{n+1}-x_n||<\text{tol}(X)$) 
and  the function tolerance  %$|f(x_{n+1})-f(x_n)|<\text{tol}F$ 
was set to tolF=0.01. The maximum number of iterations was set to 100.

\section*{Results}
\subsection*{Healthy controls}
\begin{figure}[t]
\begin{center}
\begin{picture}(400,190)
\includegraphics[width=12.5cm]{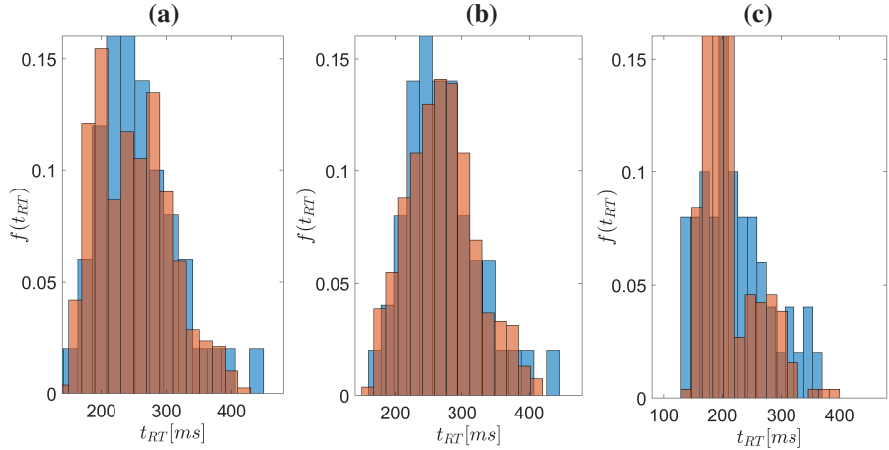}% This is a *.eps file
\put(-300,170){\textbf{(a)}}
\put(-180,170){\textbf{(b)}}
\put(-75,170){\textbf{(c)}}
\end{picture}
\end{center}
\caption{The computed distributions of reactions times (RT) for healthy controls (brown color), using the value parameters estimated by numerical optimization. The experimental distributions are also depicted for comparison purposes. \textbf{(a)} Distribution of all RT (correct antisaccades and error prosaccades): mean value $\bar{t}_{RT}=249.38$, standard deviation $s=57.14$ ($\bar{t}_{RT}=253.6$ and $s=60$ are the corresponding experimental values). \textbf{(b)} Distribution of RT only for the correct antisaccades: mean value $\bar{t}_{RT}= 269.48$, standard deviation $s=51.72$ ($\bar{t}_{RT}=270$ and $s=55$ are the corresponding experimental values).\textbf{(c)} Distribution of error prosaccades RT: mean value $\bar{t}_{RT} = 207.41$, standard deviation $s=42.76$ ($\bar{t}_{RT} =208.94$ and $s=42.59$ are the corresponding experimental values).}
\label{fig:control_exp_opt}            
\end{figure}

The numerical optimization for the healthy controls resulted to the following values of the model parameters: $r_1 = 2.135$ (95\% \text{CI}: 1.986, 2.283), $r_2 = 0.631$ (95\% \text{CI}: 0.591, 0.672) ($g_{\text{AHP}}$) and $r_3 = 1.33$ (95\% \text{CI}: 1.197, 1.47), ($g_{\text{NMDA}}$), while for the network's topology, the switching probability was found to be $p=0.01 (95\% \text{CI}: 0.007,  0.0145)$, see also tab.~\ref{tab:norm}. 
%The confidence intervals for each parameter are $r_1\in(1.986, 2.283), r_2\in (0.591, 0.672), r_3 \in (1.197, 1.47)$ and for the network probability $p\in (0.007,    0.0145)$.
The residual of the objective function was 0.0022. For the optimal parameter values, the model predicted an error rate of 34\% for the control group which was not significantly different from the experimentally observed one at a 0.05 significance level. Fig.\ref{fig:control_exp_opt} depicts the distribution function of reaction times for the experimentally observed and those derived by simulations.
%\begin{table}%[tbhp]
%\centering
%\caption{Computed values using the fminsearch in the healthy %normal case.}
%\begin{tabular}{lrrrr}
%Name of the parameter & Computed value & Estimation Interval \\
%\midrule
%1. $r_1$ & 2.28 & $r_1\in (2.23, 2.33)$ \\
%2. $r_2$  & 0.752 & $r_2\in (0.72, 0.79)$\\
%3. $r_3$ & 1.682 & $r_3\in (1.64, 1.72)$\\
%4. $p$ & 0.101 & $p \in (0.09, 0.11)$ \\
%\bottomrule
%\end{tabular}
%\addtabletext{nomenclature for the TSs refers to the numbered species in the table.}
%\end{table}
\begin{table}[h]
\centering
\begin{tabular}{ |p{4cm}||p{3cm}|p{3cm}| }
 \hline
 %\multicolumn{3}{|c|}{Country List} \\
 \hline
Parameter & Estimated values & 95\% Confidence Intervals\\
 \hline
1. $r_1$ & 2.135 & $r_1\in (2.23, 2.33)$ \\
2. $r_2$  & 0.631 & $r_2\in (0.72, 0.79)$\\
3. $r_3$ & 1.33 & $r_3\in (1.197, 1.47)$\\
4. $p$ & 0.0101 & $p \in (0.007, 0.0145)$ \\
 \hline
\end{tabular}
\newline
 \caption{Estimated values of model parameters along with their 95\% confidence intervals for the healthy controls.}
 \label{tab:norm}
\end{table}
In order to compare between the experimental and computational RT distributions, we performed a Kolomogorov-Smirnov test. The statistical test showed that the null Hypothesis that the RT data  come from the same population cannot be rejected at a 0.05 significance level (p-value: 0.77).

%\begin{figure}%[tbhp]
%\centering
%\includegraphics[width=1\linewidth]{cum_tot.eps}
%\caption{Cumulative Distribution for the Reaction Times. Blue lines correspond to experimental data and red lines are the resulting distributions that comes from the model under the optimization procedure. The first 2 higher lines correspond to normal controls case and the lines with crosses are for Schizophrenic patients.} 
%\label{fig:Distributions}
%\end{figure}
\subsection*{Patients with Schizophrenia}
\begin{figure}[t]
\begin{center}
\begin{picture}(400,190)
\includegraphics[width=12.5cm]{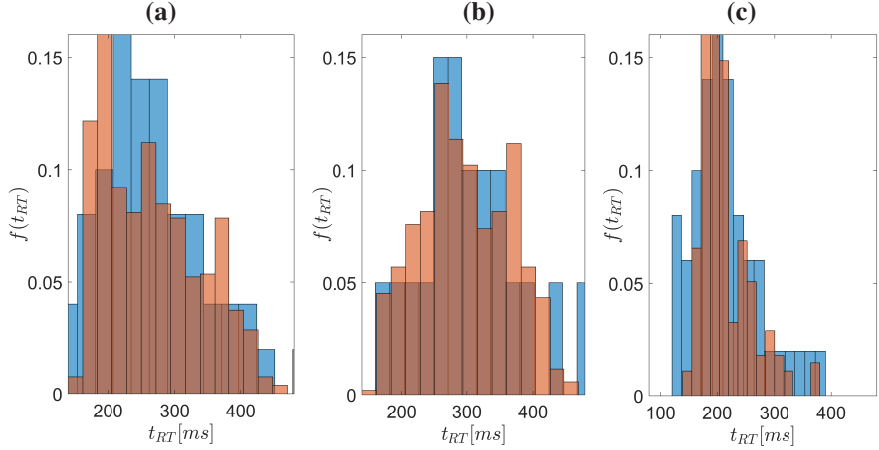}% This is a *.eps file
\put(-300,170){\textbf{(a)}}
\put(-180,170){\textbf{(b)}}
\put(-80,170){\textbf{(c)}}
\end{picture}
\end{center}
\caption{Computed distributions of reactions times (RT) for patients as derived using the parameter values estimated by numerical optimization (brown color). The experimental distributions are also depicted for comparison purposes (blue color). \textbf{(a)} Distribution of all RT (correct antisaccades and error prosaccades). Error prosaccades: mean value $\bar{t}_{RT}=266$, standard deviation $s=73.9$ (the corresponding experimental statistical moments are $\bar{t}_{RT}=266, s=81$). \textbf{(b)} Distribution of antisaccade only RT: mean values $\bar{t}_{RT}= 296.87$, standard deviation $s=70$ (the corresponding experimental statistical moments are: $\bar{t}_{RT}=304, s=77$). \textbf{(c)} Distribution of error prosaccades RT: mean value $\bar{t}_{RT}=212.63$, standard deviation $s=52$.}
\label{fig:SZ_exp_opt}            
\end{figure}

The resulting optimal values for the group of patients were: $r_1 = 2.45 (95\% \text{CI}:2.328, 2.58), r_2 = 0.885 (95\% \text{CI}:0.816, 0.954), r_3= 0.857 (95\% \text{CI}:0.805, 1.1), p=0.02 (95\% \text{CI}:0.008, 0.04)$ for the switching probability.
%The confidence intervals for each parameter are $r_1\in(2.328, 2.58), r_2\in (0.816, 0.954), r_3 \in [0.805, 1.1)$ and for the network probability $p\in(0.008, 0.04)$.
The residual of the objective function was 0.003. With these parameter values, the model predicted an error rate of 44\% for the group of patients, which was not significantly different from the experimentally observed one.  Fig.\ref{fig:SZ_exp_opt} depicts the distribution of RT antisaccades for the group of patients. Experimental and model-derived RT are in a good agreement. By applying the Kolmogorov Smirnov test, it results that the $H_0$ hypothesis that the two sets of data come from the same distribution cannot be rejected at the 0.05 significance level (p-value: 0.543).

% \begin{figure}%[tbhp]
% \centering
% \includegraphics[width=.8\linewidth]{schizoAntisac.eps}
% \caption{Small world network. Staring with lattice like structure with add remote connections with probability p.}
% \label{fig:experiment}
% \end{figure}

\begin{figure}[ht!]
\begin{center}
\begin{picture}(400,190)
\includegraphics[width=12.9cm]{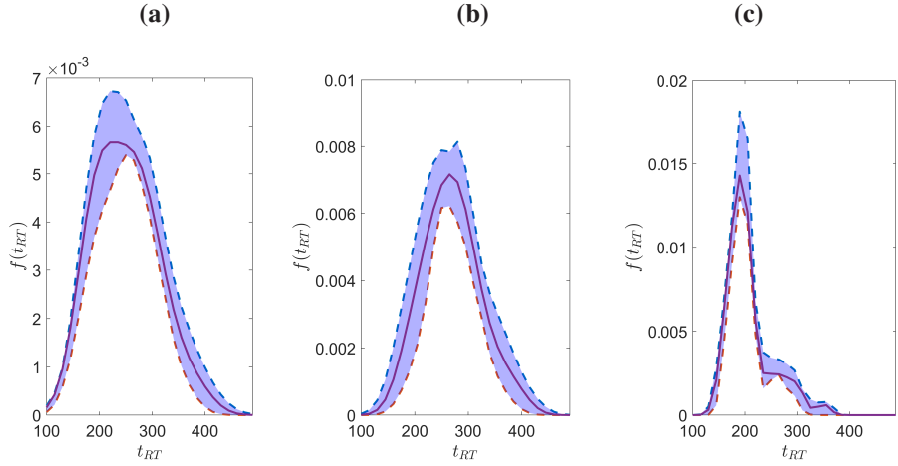}% This is a *.eps file
\put(-300,170){\textbf{(a)}}
\put(-180,170){\textbf{(b)}}
\put(-75,170){\textbf{(c)}}
\end{picture}
\end{center}
\caption{Healthy controls. The computed distributions of reaction times (solid line); the shaded area depicts the 95\% confidence intervals. \textbf{(a)} Distribution of all reaction times (RT) (correct antisaccades and wrong prosaccades). \textbf{(b)} Distribution of antisaccades only RT. {(c)} Distribution of error prosaccades RT.}
\label{fig:healthy_95per}            
\end{figure}

\begin{figure}[ht!]
\begin{center}
\begin{picture}(400,190)
\includegraphics[width=12.5cm]{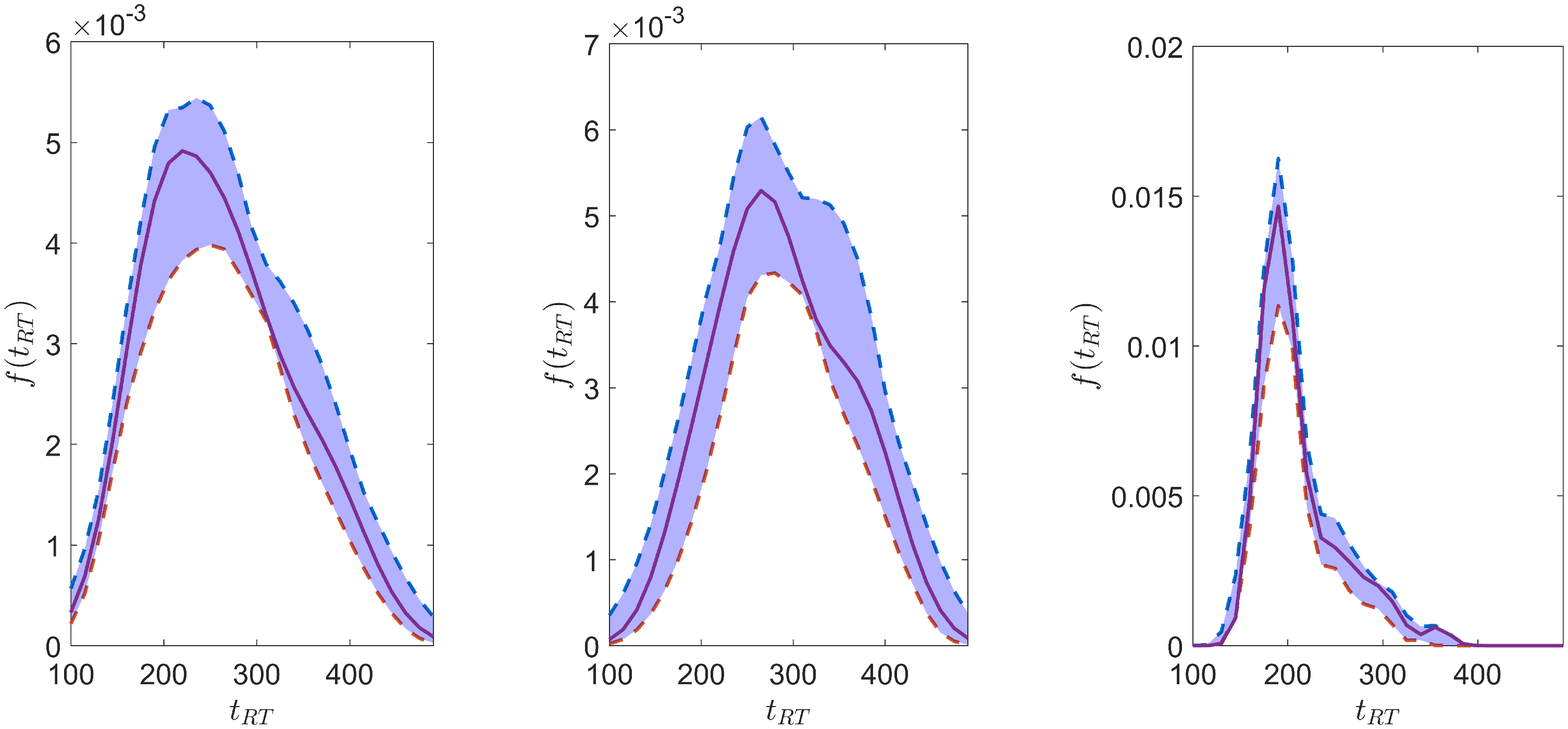}% This is a *.eps file
\put(-300,170){\textbf{(a)}}
\put(-180,170){\textbf{(b)}}
\put(-75,170){\textbf{(c)}}
\end{picture}
\end{center}
\caption{Schizophrenic patients. Computed distributions of reaction times (solid line); the shaded area depicts the 95\% confidence intervals. \textbf{(a)} Distribution of all reaction times (RT) (correct antisaccades and wrong prosaccades). \textbf{(b)} Distribution of antisccade only RT. {(c)} Distribution of error prosaccades RT.}
\label{fig:schizo95_per}            
\end{figure}

\begin{figure}[t]
\centering
\begin{picture}(400,270)
\includegraphics[width=.98\linewidth]{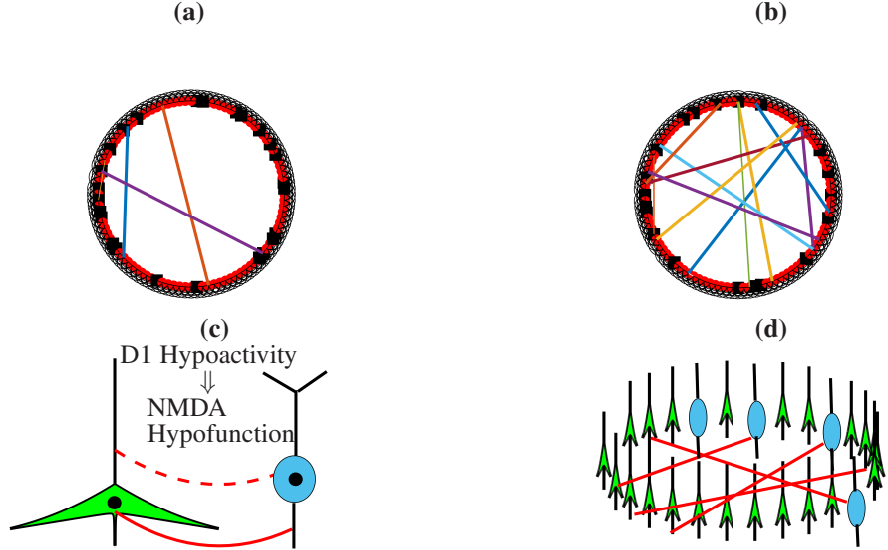}
\put(-340,240){\textbf{(a)}}
\put(-120,240){\textbf{(b)}}
\put(-330,120){\textbf{(c)}}
\put(-120,120){\textbf{(d)}}
\put(-360,110){D1 Hypoactivity}
\put(-330,100){$\Downarrow$}
\put(-350,90){NMDA }
\put(-350,80){Hypofunction}
\end{picture}
\caption{A schematic of the PFC neural network. \textbf{(a)} A low number of remote connections is characteristic of the topology for the group of healthy controls. \textbf{(b)} A more random topology, i.e. an increased number of remote connections, resulting from larger values of the switching probability in the Watts and Strogatz \cite{Wat98,Bull09} model is characteristic for the group of schizophrenic patients. \textbf{(c)} A schematic of a disrupted  circuit model in patients with schizophrenia. The circuit consists of cortical pyramidal neurons (green) and GABAergic neurons. A D1 hypofunction implies NMDA receptor hypofunction which results to a hypoactivity of the cortical pyramidal neurons. Due to this, the communication between pyramidal and interneurons is disrupted. \textbf{(d)} A schematic of the PFC network; an abnormal behavior of the pyramidal-interneuron circuit in conjunction with the network disruption (randomization) leads to ineffective performance of the PFC which is linked to the SC network.} 
\label{fig:Mechanism}
\end{figure}

Figs.~\ref{fig:healthy_95per} and \ref{fig:schizo95_per} depict the mean and 95\% confidence interval estimations of the RT. Fig.~\ref{fig:Mechanism} shows a schematic of two characteristic PFC neural networks for the healthy controls (Fig.\ref{fig:Mechanism}(a)) and for the group of patients (Fig.\ref{fig:Mechanism} (b)); in the second case the resulting PFC network is found to be more random with an increased number of random remote connections. It has been suggested that such random structures act as communication-integration disruptors \cite{Hig07,Lo15}. In concurrent, Fig.\ref{fig:Mechanism}(c) 
depicts a schematic of a local circuit between a pyramidal excitatory PFC neuron and a GABAergic interneuron (microscopic scale). The optimization process manifested here a mechanism by which the  dopamine D1 results to a hypofunction of NMDA neurons (lower $g_{\text{NMDA}}$ in patient case) disrupting this way the functionality of NMDA-GABAergic dipoles. In the higher level of  the neural PFC network (mesoscopic scale) (see  Fig.\ref{fig:Mechanism}(d)), the abnormal behavior of the pyramidal-interneuron circuit together with the network disruption leads to an ineffective behaviour performance at the macroscopic level.

\begin{table}[t]
\centering
\begin{tabular}{ |p{4cm}||p{3cm}|p{3cm}| }
 \hline
 %\multicolumn{3}{|c|}{Country List} \\
 \hline
 
Parameter & Estimated value & 95\% Confidence Intervals\\
 \hline
1. $r_1$ & 2.456 & $r_1\in ( 2.328, 2.584)$ \\
2. $r_2$  & 0.885 & $r_2\in (0.816, 0.954)$\\
3. $r_3$ & 0.957 & $r_3\in (0.805, 1.11)$\\
4. $p$ & 0.02 & $p \in (0.0075, 0.04)$ \\
 \hline
\end{tabular}
\newline
 \caption{Estimated values of model parameters along with their 95\% confidence intervals for the group of patients with schizophrenia.}
  \label{tab:schiz}
\end{table}

\section*{Discussion}
Numerical simulations of the proposed PFC-SC biophysical-based network model that was developed to assess the antisaccade performance in healthy controls and patients with schizophrenia predicted the hypothesis about the effects of the changes in the dopamine levels, NMDA functionality and simultaneously the  alteration of the PFC network topology. In particular, decreased DA levels in the PFC, combined with the NMDA hypofunction and the connectivity topology alteration predicted all basic antisaccade performance differences as have been observed experimentally in patients with schizophrenia (see introduction), namely a large increase in the error rate as well as an increase in the mean and variance of the reaction times for the correct antisaccades. 
%\begin{table}%[tbhp]
%\centering
%\caption{Computed values using the fminsearch in the Schizophrenic Patient case}
%\begin{tabular}{lrrrr}
%Name of the parameter & Computed value & Estimation Interval \\
%\midrule
%1. $r_1$ & 2.874 & $r_1\in (2.859, 2.89)$ \\
%2. $r_2$  & 1.113 & $r_2\in (1.108, 1.135)$\\
%3. $r_3$ & 1.472 & $r_3\in (1.465, 1.487)$\\
%4. $p$ & 0.14 & $p \in (0.136, 0.149)$ \\
%\bottomrule
%\end{tabular}
%\addtabletext{nomenclature for the TSs refers to the numbered species in the table.}
%\end{table}
Here, it is important to stress that the numerical optimization for the calibration of the model parameters was performed based on the total experimental distribution which contains both antisaccade and error prosaccade RT. Simulation results were in a very good agreement with the experimentally observed ones (both antisaccades and error prosaccades), thus providing evidence that the proposed model is able to capture the essential aspects of the observed data.\par
The estimated values of the  model parameters for the group of patients suggest a deviant in the PFC activity. More specifically, the conductance of the inhibitory current $I_{\text{AHP}}$ current was found higher in the group of patients (2.456 in patients vs. 2.135 in healthy controls) suggesting an inhibition on the PFC neurons. The increment of $I_{\text{AHP}}$ decreases the firing rates of the PFC neurons which in turn results to a lower value of the $slope_p$ (see Fig.~\ref{fig:Simulations}(b)). This suggests that the buildup neurons trigger more slowly, i.e. they delay to reach the critical value that governs the RT, thus leading to larger reaction times. This result is consistent with the dopamine hypo-activity hypothesis \cite{Yi13, Malenka86, Ped95}. In \cite{Yi13}, using cell patch clamp recording in rat prefrontal neurons showed that the DA activation (under the D1 receptor) suppress the $I_{\text{AHP}}$ current, thus leading to a higher spiking activity. Pezedrani et al.\cite{Ped95} also reports  similar results regarding the Hippocampus neurons.\par
The conductance of the excitatory synaptic current $I_{\text{NMDA}}$ for the group of patients was found smaller with respect to the one for the group of healthy control (0.957 in patients vs 1.33 in healthy controls), which is consistent with the glutamate hypofunction theory. The reduction of the NMDA conductance weakens the current transmission, impairs the  neurons interactions, thus resulting to smaller firing rates of the PFC neurons. The decreased PFC  activity weakens the planned input to SC (see sec.\ref{sec:SC}). As a consequence, the planned antisaccade processing that would be appropriate for the given task instruction is defective and the SC produces longer and more variant reaction times.
%This can be also explained by the dopamine effect  acting on pyramidal cells located in layers V and VI of the medial prefrontal cortex of rats, concluded that dopamine at low concentrations acts preferentially on D1-like receptors to enhanced .Similar results are produced from Zhang et al. [PLos one] on Nucleus accumbens rat neurons, concluding that the effect of DA on evoked post synaptic current ( ) were mimicked by the D1-like receptor agonist SKF 38393 and lead to 90 per cent reduction of. 
Furthermore, the results of our study coincide with the disconnetivity  hypothesis, that for the pathological case one would expect a disrupted neuronal-functional connectivity in the the PFC-SC network. In our model, the PFC was modeled by leaky integrate-and-fire dynamics evolving on a small-world network structure. In the case of the healthy control group, the value of the switching probability that determines the topology of the small-world network  dictates a high clustered (similar to a ring) structure. Instead, in the group of patients, the switching probability was higher: $p \sim 0.01$ vs $p \sim 0.02$ for the group of healthy-control dictating a connectivity that is closer to a random structure. In the review article \cite{Hao07} (and the references therein), the authors report that under working memory tasks, schizophrenic patients involve a sparser network of cortical regions resulting to a reduced cortical signal to noise ratio (which can be interpreted as a loss of signal information). Disruptions in connectivity were also  reported in auditory  hallucinations, compared with healthy controls and patients who had schizophrenia but not hallucinations \cite{Hubl04,Day16}, suggesting that the auditory hallucinations originate from altered connectivity of the same regions that process normal hearing and speech \cite{Hig07}. 
Similar results have been reported by Rubinov et al. \cite{Rub09} who studied two groups (healthy controls and patients with schizophrenia) with resting state EEG. The authors concluded that cognitive disturbances may be due to the randomization of the underlying functional connectivity network.  In Lo et al. \cite{Lo15} it is also reported significant randomization of global network metrics for patients with schizophrenia. These conclusions are in line with our model outcomes as changes in the network topology going from clustered and structured to more random ones result in a worse cognitive performance in the antisaccade task.\par
Based one the above, our proposed model and numerical results may  provide useful insights regarding the organization of the connectome and its alteration, especially on the hypothesis of the ``small-worldness" of key parts of the sensorimotor system (in particular of the PFC) and the study of its topological characteristics. Even though, functional neuroimaging studies have provided evidence of the small-world organization of the brain \cite{Bull09}, the connectome and its plasticity within and between areas of the cortex such as the PFC that possess a prominent role in the voluntary control of movements including saccades and the midbrain such as the SC is still not well understood \cite{Meer12}. Several neuroimaging studies that have focused toward this direction have revealed that there are certain differences in the functional connectivity of the PFC between healthy controls and subjects suffering from severe mental disorders such schizophrenia \cite{Yang12, Yu13,Gao2021}. The outcomes of our model was in line with the main findings of these studies.\par 
In conclusion, our PFC-SC biophysical network model for the approximation and assessment of the reaction times of the antisaccade task showed that the combination of three factors, namely the DA and NMDA hypofunction and the network alteration in the PFC explain the deficits in the antisaccade performance as observed experimentally in patients with schizophrenia. Furthermore, the connectivity changes, as the disconnection hypothesis suggests in the PFC connectome resemble (qualitatively) those observed in functional connectivity neuroimaging studies.

%Without assuming any prescribed network topology for the PFC, as this could be principally vary between lattice to regular random connectivity, the black-box optimization outcomes resulted to higher values of clustering and local efficiency in the heatlhy controls when compared with the patients with schizophrenia.    
Another direction as to which the proposed model could be exploited in future studies is the investigation of the so-called Stochastic Resonance mechanism \cite{Wies95, Kita06, Mend12}, which is related to improved reaction performance in the presence of a low level background noise within a certain interval. Future research could be also targeted on the study of the potential effects of DA-D2 receptors on the neuronal responses of pyramidal neurons in the cortical modules and subsequently on the behavioral data of the patient group. D2 receptors tend to have opposing effects relative to the action of the D1 component, as far as the enhancement or inhibition of cortical currents is concerned \cite{Seam04}. In addition, the  extension of the model to include  the cortico-striatal, the basal ganglia network constitutes another future direction. Studies in neurological inhibitory control, suggest that inhibition is achieved via a fronto-basal-ganglia network, which could intercept the Go process and thus decrease thalamocortical output \cite{Aron07,Wes16,Hest20}. The augmented model will be used to simulate the control override and inhibition of the habitual response selection mechanism. \cite{Wiec13,Coe19,Aron07}.

   \section*{Acknowledgements}
C.S. acknowledges partial support by INdAM, through GNCS research projects. K.S. and J.S. thank the DFG for support through the Collaborative Research Center CRC 1270 (Deutsche Forschungsgemeinschaft, Grant/ Award Number: SFB 1270/1–299150580).

\end{document}